# Compact Modeling of MOSFET I-V Characteristics and Simulation of Dose-Dependent Drain Currents

G. I. Zebrev, V. V. Orlov, A. S. Bakerenkov, V. A. Felitsyn

*Abstract* — We have presented a compact MOSFET model, which allows us to describe the I-V characteristics of irradiated long-channel and short-channel transistors in all operation modes at different measurement temperatures and interface trap densities. The model allows simulating of the off-state and the on-state drain currents of irradiated MOSFETs based on an equal footing. Particularly, a novel compact model of the rebound effect in the n-MOSFETs was employed for simulation of the total dose dependencies of drain currents in the highly scaled 60 nm node circuits irradiated up to 1Grad. Compatibility of the model parameter set with BSIM and a single closed form of the model equation imply the possibility of its easy implementation into the standard CAD tools.

## I. Introduction

Electrical level simulation of digital circuits is essentially based on compact models of MOSFETs [1, 2]. Models of circuit elements should be sufficiently simple and accurate to be incorporated in circuit simulators. These models are needed to numerically compute the device characteristics accurate and fast enough to simulate electrical circuits. The contradictory objectives of model simplicity and accuracy make the compact modeling the challenging research area jointly in device physics, electronic engineering and applied mathematics [3]. A condition of fast simulation implies an analytic form of the compact models. Radiation hardness and operation in harsh environments impose the additional requirements for accuracy of MOSFET compact models. As noted in [4]: "Understanding and analyzing the impact of ionizing radiation and aging effects in modern MOS technologies requires the incorporation of radiation and stress-induced defects into advanced compact model formulations of device operation." In particular, it would be desirable the model would provide an opportunity to accurately describe the MOSFET current-voltage (I-V) characteristics at different doses of ionizing irradiation over a wide temperature range.

Irradiation impacts mainly through the buildup of the oxide traps and rechargeable interface traps, which influence on the threshold voltage $V_T$, the subthreshold slope *SS*, the low-field mobility $\mu_0$, etc. It should be noted that the charge trapping and interface trap buildup in the gate oxide are negligible at moderate doses (< 1 Mrad) in the highly scaled CMOS circuits due to a very low thickness of the gate oxides. Nevertheless, the problem of radiation-induced drain current degradation in the above threshold and the subthreshold operation modes remains valid even for the short-channel MOSFETs with the technology nodes 65 and 130 nm, in particular, in the Application-Specific Integrated Circuits (ASICs) irradiated up to a total dose level ~1 Grad [5]. Many industrial and space applications of the CMOS circuits imply a combined impact of irradiation and permanent temperature cycling. This makes necessary to study radiation behavior of MOSFETs as a function of operation temperature. A consistent modeling and simulation of the entire I-V curve over the typical current range $10^{-9}$ – $10^{-3}$ A for a wide temperature range and different irradiation conditions represents a challenging task. This is particularly relevant for the calculation of radiation-induced leakage through the parasitic edge transistors [4, 6] and for simulation of microdose induced drain leakage effects in the power trench MOSFETs [7]. The objective of this paper is to demonstrate the ability of the proposed model to accurately simulate the MOSFET I-V characteristics at different temperatures and the dose-dependent drain currents.

The rest of this paper is organized as follows. The Sec. II focuses on a concise description of the physics-based MOSFET model and its features. The model validation and radiation-oriented applications are presented in Sec. III and IV.

## II. MOSFET's I-V Model Formulation

### A. Historical background and the distinctive features of the model

Early compact MOSFET models were relied on the explicit dependencies of the drain current $I_D$ as functions of gate voltage $V_G$ and drain voltage $V_D$ with the phenomenological threshold voltage $V_T$ as a parameter [8]. Such approach omits relevant physics and it is not able to describe the effect of interface traps density on the subthreshold region of I-V characteristics. Most of the MOSFET models have a piecewise structure, where the separate equations are utilized to simulate the different operation regions. This is linked with a fine question of diffusion and drift currents in the subthreshold and above threshold regions of I-V characteristics. A piecewise approximation complicates the description of I-V characteristics in the intermediate region between the sub-threshold and the above threshold operation modes, which is critical for radiation-oriented applications. Despite their limitations, the threshold-voltage-based models (e.g. BSIM4 [9]) have been successfully used for many circuit design work. The surface-potential-based models [10] are more physical and more appropriate for radiation applications. This class of models requires a numerical solution of a non-linear equation relating the external gate voltage and internal surface potential $V_G(\varphi_S, params)$ to obtain the inverse relation $\varphi_S(V_G, params)$, followed by computation of the channel carrier concentration $n_S[\varphi_S(V_G)]$ and the drain current $I_D[\varphi_S(V_G)]$.

For example, Sanchez Esqueda et al. [4] proposed to calculate the drain current and charges in MOSFETs through the Pao-Sah double integral formula [11], or, using the charge-sheet approximation [12]. Such approach is a computationally-intensive one, and it does not belong, in fact, to a class of analytical compact models. Moreover, the calculation of the internal surface potential $\varphi_S$ is an intermediate step, which, in fact, is not necessary to simulate the drain current. We have formulated a version of the diffusion-drift MOSFET model, where the drain current is calculated as an explicit function of $n_S$ instead of $\varphi_S$. The diffusion to drift current ratio is also modeled as a function of $n_S$, allowing us to simulate the entire MOSFET I-V characteristics at different temperatures and interface trap densities. This makes it possible to calculate the dose response of transistors over the entire range of the drain currents and doses, based on a single compact formula with a restricted set of physical parameters.

*B. The model description*

We use in this paper the diffusion-drift MOSFET model, initially proposed in [13, 14]. This model relies on an explicit solution of the channel current continuity equation, and it has been consistently implemented for different types of the field-effect devices including SOI and double-gate transistors [15], graphene FETs [16, 17], and molybdenite MoS$_2$ monolayer transistor [18]. This model is able to describe the entire I-V curves via a single analytic expression both in the subthreshold and above threshold regions, as well as in the linear and saturation modes.

An original explicit form of drain current $I_D$ for an ideally well-designed (i.e., without any geometrical short-channel effects) MOSFET has been written as follows

$$I_D = I_{DSAT}\left(1-\exp\left(-2\frac{V_D}{V_{DSAT}}\right)\right), \quad (1)$$

where $I_{DSAT}$ is a saturation current, $V_D$ is the drain-source bias. The drain saturation voltage $V_{DSAT}$ is an explicit function of the mobile charge density

$$V_{DSAT} \cong \varphi_T\left(1+\frac{C_{it}}{C_{ox}+C_D}\right)+\frac{qn_{S0}}{C_{ox}+C_D}, \quad (2)$$

where $q$ is the electron charge, $n_{S0}$ is the channel carrier density near the source, $\varphi_T = k_BT/q$ is the thermal potential, $C_{ox}$, $C_D$, $C_{it}$ are the oxide, the depletion layer and the interface trap capacitance per unit area, correspondingly. The MOSFET electrostatic saturation current $I_{DSAT}$ is represented as a sum of the diffusion and the drift components

$$I_{DSAT} = \frac{W}{L}qD_0n_{S0} + \frac{W}{L}\frac{\mu_0 q^2 n_{S0}^2}{2(C_{ox}+C_D)}, \quad (3)$$

where $W$ and $L$ are the channel's width and length, $q$ is the electron charge, $\mu_0$ and $D_0$ are the channel carrier mobility and diffusivity, coupled by the Einstein relation $D_0 = \mu_0\varphi_T$. The first term in (3), corresponding to a linear dependence of the drain current on $n_{S0}$, is a diffusion current, dominating in the subthreshold region at low $n_{S0}$. The second term in (3) corresponds to the saturated drift current in the above-threshold mode in a well-known square-law approximation [19].

An accurate simulation of the channel charge density over 6-7 orders of its magnitude is a difficult task even for a simple planar geometry of the bulk MOSFETs. Therefore, it is more convenient to model $n_{S0}(V_G)$ by a phenomenological interpolation which is similar to that used in the CMOS design compact model BSIM [9]

$$qn_{S0} = \frac{2C_{ox}m\varphi_T \ln\left[1+\exp\left(\frac{V_G-V_T}{2m\varphi_T}\right)\right]}{1+2m\frac{C_{ox}}{C_D}\exp\left(-\frac{V_G-V_T}{2m\varphi_T}\right)}, \quad (4)$$

where $m = 1+(C_{it}+C_D)/C_{ox}$ is the ideality factor which is closely related to the logarithmic subthreshold slope $SS = m\varphi_T \ln 10$ [20]. This interpolation is validated by its asymptotes in the above-threshold ($qn_{S0} \cong C_{ox}(V_G-V_T)$) and the subthreshold $qn_{S0} \cong C_D \varphi_T \exp[(V_G-V_T)/m\varphi_T]$ regions. The threshold voltage $V_T$ and ideality factor $m$ are the radiation-sensitive parameters in (4).

*C. Short channel effects*

The carrier speed saturation is an inherent short-channel effect [21]. To take it into account, equation (1) should be modified as follows [17]

$$I_D = \frac{W}{2L}q\mu_0 n_{S0}\tilde{V}_{DSAT}\left(1-\exp\left(-2\frac{V_D}{\tilde{V}_{DSAT}}\right)\right). \quad (5)$$

Here, a generalized saturation voltage $\tilde{V}_{DSAT}$ is defined as

$$\tilde{V}_{DSAT} = \frac{2v_{max}L}{\mu_0}\tanh\left(\frac{\mu_0 V_{DSAT}}{2v_{max}L}\right), \quad (6)$$

where $v_{max}$ is a maximum carrier speed (~ $10^7$ cm/s in silicon), which is saturated to due to the optical phonon emission. For the long-channel devices, when the argument of the hyperbolic tangent is much less than a unity, we have $\tilde{V}_{DSAT} \cong V_{DSAT}$, corresponding to the square-law approximation $I_{DSAT} \propto (V_G - V_T)^2$.

Otherwise, i.e. for the short-channel MOSFETs and/or for the large gate voltages, we have an approximation $\tilde{V}_{DSAT} \cong 2v_{max}L/\mu_0$, and, accordingly,

$$I_{DSAT} \cong Wqn_{S0}v_{max}, \quad (7)$$

corresponding to $I_{DSAT} \propto (V_G - V_T)$. In practice, the saturation current of the silicon MOSFET increases approximately as $(V_G - V_T)^\alpha$, where $1 \leq \alpha \leq 2$.

### D. Impact of interface traps at different temperatures

The channel carrier density is affected by the oxide trapped charge and interface traps via the threshold voltage and subthreshold swing *SS*, which is bound with the interface trap capacitance $C_{it} = q^2 D_{it}$. We have chosen $V_T$ and the subthreshold swing SS as independent variables for the description of I-V curves. The carrier mobility at room and elevated temperatures $T$ is mainly determined by the phonon scattering, and it can be simulated as to be proportional to $T^{-3/2}$ [22]. The channel carrier mobility generally is also the vertical electric field dependent function due to surface roughness scattering. For simplicity, the dependence on the gate voltage can be modeled by an empirical relation

$$\mu_0(T, V_G) = \frac{\mu_0}{1 + \theta q n_S / C_{ox}} \left(\frac{T_0}{T}\right)^{3/2}, \quad (8)$$

where $T_0$ is a reference temperature (assumed to be equal 300 K), $\theta$ is a constant which is assumed here to be dose-independent. The threshold voltage of n-MOSFETs generally decreases with temperature due to the Fermi level temperature change in the Si bandgap [21]

$$V_T(T) = V_{T0} - \alpha_{VT}(T - T_0), \quad (9)$$

where $\alpha_{VT}$ is the threshold voltage temperature coefficient.

## III. I-V MODEL VALIDATION

Figures 1(a) and 1(b) show experimental and simulated drain currents of the unirradiated long-channel n-MOSFETs measured as functions of the gate voltage at different temperatures [23]. A set of experimental I-V characteristics at different measurement temperatures exhibits a pretty typical common point of intersection for not very wide temperature range. The origin of this point has no fundamental reasons and can be explained by the temperature-dependent shift of threshold voltage and decreasing in mobility [24, 25]. The model constants $V_T$, $C_{it}$, $\mu_0$, $\theta$, and, in fact, $C_D$, are fitted to describe the I-V characteristics before and after irradiation at a fixed temperature. The parameter $\alpha_{VT}$ is used to describe I-V curves at different temperatures.

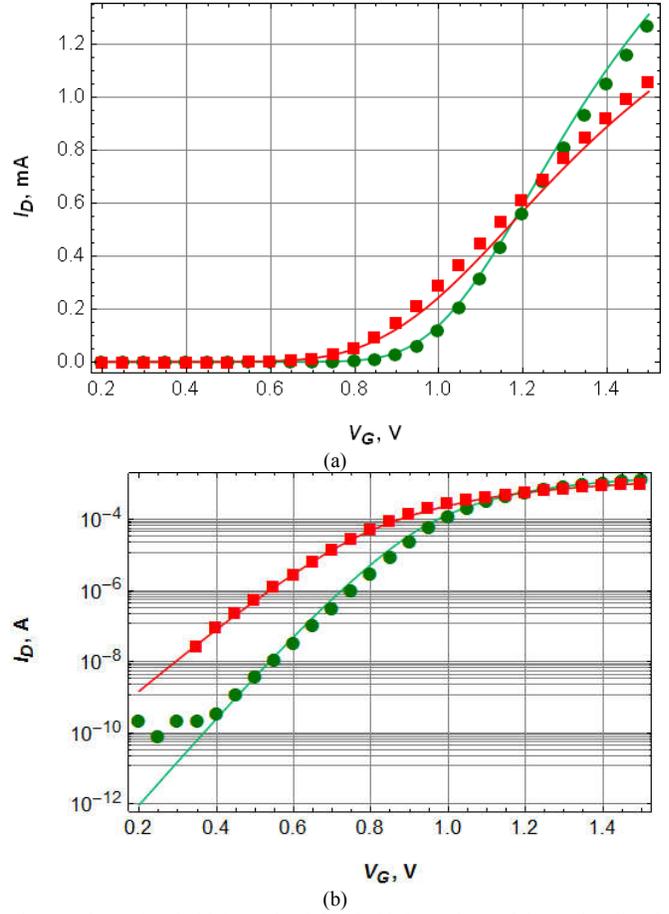

Fig. 1. Above threshold (a) and subthreshold (b) regions of the un-irradiated long-channel MOSFET's transfer characteristics measured at different temperatures 28 °C (circles) and 125 °C (squares), W/L = 1000μm/1μm, $t_{ox}$ = 20 nm, $C_D/C_{ox}$ = 0.3. Fitted parameters are $V_T$ = 0.7 V, $\mu_0$ = 275 cm$^2$/(V×s), $D_{it}$ = 1.5×10$^{12}$ cm$^{-2}$eV$^{-1}$, $\alpha_{VT}$ = 2 mV/K, $\theta$ = 0.35 V$^{-1}$.

Figure 1 shows the comparison of simulation and experimental I-V curves at different temperatures both in the above-threshold and the subthreshold regions. In fact, measurements in Fig. 1 were performed for n-MOSFET which is a part of CMOS inverter at a fixed $V_{OUT} = V_{DS}$ = 50 mV and $V_{GS} = V_{IN} = V_{DD}$.

Figure 2 shows a set of I-V curves of the same MOSFETs at different temperatures, after multiple irradiations and annealing cycles. Irradiations were performed with the all grounded terminals at the x-ray source with a total dose of order a few Mrads (Si). Multiple irradiations followed by annealing at elevated temperature (~ 120 °C) result in a slight shift of the threshold voltage towards the positive direction, which corresponds to sufficiently annealed oxide-trapped charge and buildup of a significant amount of the acceptor-like interface traps (so-called the "rebound" effect [26]). The fitted parameters $V_T$, $C_{it}$, $\alpha_{VT}$ and $\mu_0$ have changed with irradiation. Interface trap capacitance after irradiation is uniquely fitted from the subthreshold part of I-V characteristics.

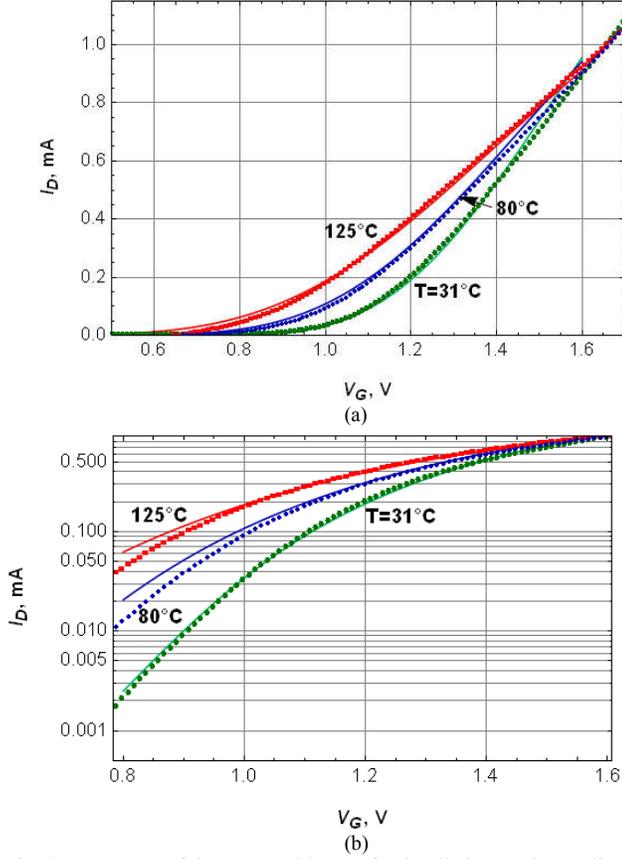

Fig. 2. I-V curves of the same MOSFET after irradiations and annealing cycles in the above threshold (a) and subthreshold (b) regions. Fitted dose-sensitive parameters are $V_{T0} = 0.7$ V, $\alpha_{VT} = 4$ mV/K, $\mu_0 = 234$ cm$^2$/(V×s), $D_{it} = 3 \times 10^{12}$ cm$^{-2}$eV$^{-1}$; $\theta = 0.35$ V$^{-1}$.

The carrier mobility $\mu_0$ is slightly decreased after irradiation and thermal stresses that correspond to additional carrier scattering on the charged interface defects. The threshold voltage temperature coefficient $\alpha_{VT}$ increases with irradiation due to an increase in interface trap density of states. Essentially, that several parameters extracted from the subthreshold and the above-threshold parts of the I-V curves are sufficient to accurately simulate an intermediate region of current-voltage characteristics at different temperatures even for irradiated devices.

The model reported above, allows us to describe I-V characteristics of contemporary short-channel MOSFETs. For instance, Fig. 3 shows a comparison of experimental (adapted from [5]) and simulated I-V curves of 60-nm MOSFETs before and after irradiation with a dose ~1 Grad.

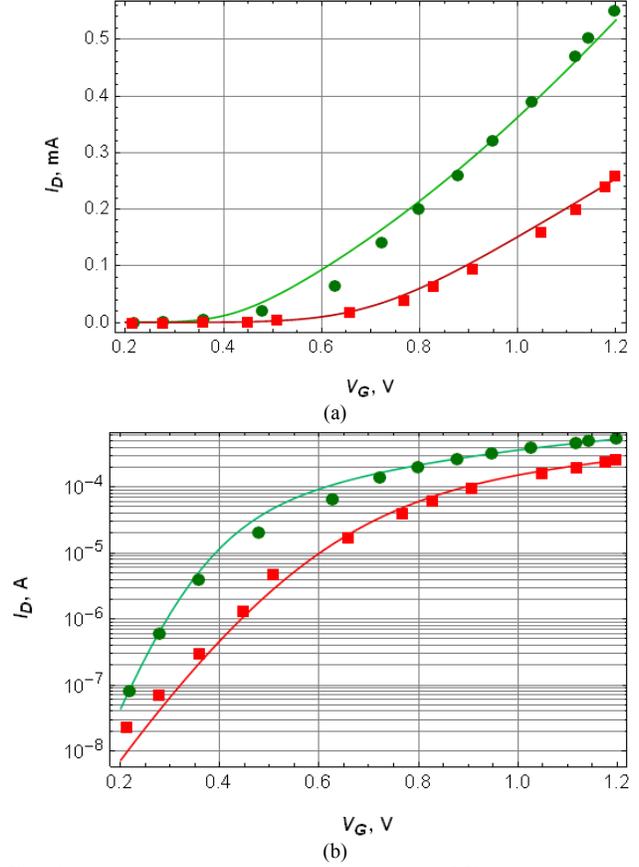

Fig. 3. Above threshold (a) and subthreshold (b) I-V curves of the short-channel n-MOSFET measured at room temperature and $V_{DS} = 1.2$ V before (circles) and after (squares) irradiation with a dose ~ 1 Grad(Si), W/L = 1 μm /60 nm, $t_{ox} = 3$ nm, $C_D/C_{ox} = 0.3$. Fitted parameters before irradiation are $V_{T0} = 0.17$ V, $\mu_0 = 81$ cm$^2$/(V c), $D_{it} = 2 \times 10^{12}$ cm$^{-2}$eV$^{-1}$, $\theta = 0$, $v_{max} = 10^7$ cm/s. The fitted values of the dose sensitive parameters after irradiation are $\mu_0 = 40$ cm$^2$/(V×s), $V_{T0} = 0.2$ V, $D_{it} = 1.1 \times 10^{13}$ cm$^{-2}$eV$^{-1}$.

Importantly, that in contrast to a previous case, the I-V characteristics in Fig.3 were measured in the velocity saturated mode, i.e. when (7) turns out to be more essential.

Figures 4-5 show the model validation for another MOSFET type.

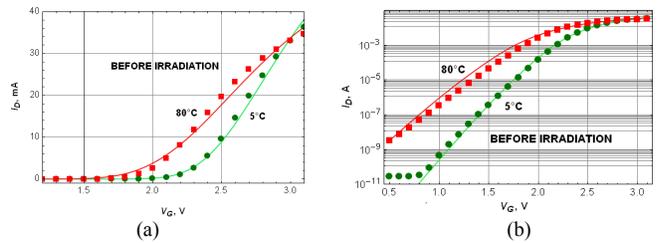

Fig. 4. Pre-irradiation above threshold (a) and subthreshold (b) I-V curves of the power n-MOSFET measured at $V_{DS} = 0.1$ V for 5 °C (circles) and 80 °C (squares). Fitted parameters before irradiation are $V_{T0} = 1.52$ V, transconductance $g_{m0} = (W/L)\mu_0 C_{ox} = 0.388$ A/V$^2$, $D_{it} = 4.5 \times 10^{12}$ cm$^{-2}$eV$^{-1}$ (at $t_{ox} = 20$ nm), $\theta = 0.2$, $\alpha_{VT} = 7.0$ mV/K.

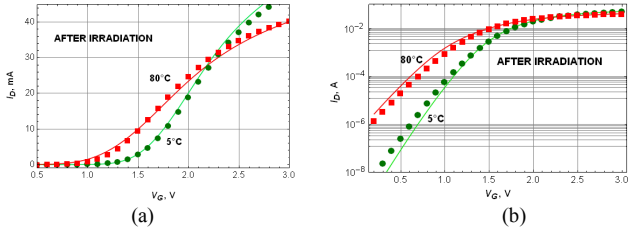

Fig. 5. Post-irradiation above threshold (a) and subthreshold (b) I-V curves of the power n-MOSFET measured at $V_{DS}$ = 0.1 V for 5 °C (circles) and 80 °C (squares). Fitted parameters after irradiation are $V_{T0}$ = 0.64 V, transconductance $g_{m0} = (W/L)\mu_0 C_{ox}$ = 0.374 A/V$^2$, $D_{it}$ = 5 ×10$^{12}$ cm$^{-2}$eV$^{-1}$, $\theta$ = 0.32, $\alpha_{VT}$ = 7.4 mV/K.

The n-channel power HEXFETs 2N7002 (up to 400mA) were irradiated at $V_G$ = 1.7V by X-rays with a dose rate ~10 rad/s up to a total dose ~10 krad at a room temperature (25 °C). After the irradiation, the transistors were annealed for 2 hours at 85 °C to eliminate the effects of post-irradiation instability. Although the examined transistors have been produced by a specific DMOS technology, a comparison of simulated and measured I-V curves demonstrates a good agreement at different temperatures before and after irradiation.

## IV. STATIC OFF-STATE LEAKAGE CURRENT MODELING

### A. Off-state drain current of n-MOSFET

Both the off-state drain current $I_{off} = I_D(V_G = 0, V_D = V_{DD})$, controlling the static leakage, and the on-state current $I_{on} = I_D(V_G = V_D = V_{DD})$, responsible for the circuit speed, are determined at the maximum supply voltage $V_{DD}$ at the drain, corresponding to a saturation mode $I_{DSAT}$.

Let us consider more the off-state current simulation features. The channel concentration at $n_{S0}^{(0)} = n_{S0}(V_G = 0)$ can be immediately derived from (4) as follows

$$qn_{S0}^{(0)} = \frac{2C_{ox} m\varphi_T \ln\left[1 + \exp\left(-\frac{V_T}{2m\varphi_T}\right)\right]}{1 + 2m\frac{C_{ox}}{C_D}\exp\left(\frac{V_T}{2m\varphi_T}\right)}. \quad (10)$$

The off-state channel concentration $n_{S0}^{(0)}$ is controlled by the dose dependence of the threshold voltage.

$$qn_{S0}^{(0)}(D) \cong \begin{cases} C_D \varphi_T \exp\left(-\frac{V_T(D)}{m\varphi_T}\right), & V_T > 0, \\ -C_{ox} V_T(D), & V_T < 0, \end{cases} \quad (11)$$

where the dose-dependent ideality factor is defined as

$$m(D) = 1 + \left(q^2 D_{it}(D) + C_D\right)/C_{ox}. \quad (12)$$

These equations illustrate the fact that the subthreshold radiation-induced leakage current is extremely sensitive to instabilities in threshold voltage and interface trap density.

The off-state leakage current both for the long-channel and short-channel MOSFETs can be calculated as the saturation drain current in (5). If the saturation voltage $V_{DSAT}$ at $V_G = 0$ is small, the off-state leakage remains unaffected by the carrier velocity saturation effects, and it can be evaluated via (3) with $n_{S0} = n_{S0}^{(0)}$. Until the threshold voltage remains positive $V_T > 0$ (e.g. at low doses or, at the rebound effect condition), the diffusion term in (3) dominates

$$I_{off}(D) \cong \frac{W}{L} q\mu_0 \left(\frac{T_0}{T}\right)^{3/2} C_D \varphi_T^2 \exp\left(-\frac{V_T(D)}{m(D)\varphi_T}\right), \quad (13)$$

Notice that the leakage current increases with temperature due to the Arrhenius form of the temperature dependence in this case. The leakage current is very sensitive to dose in this regime.

The case $V_T(D) < 0$ is also possible, e.g., at high doses with a small amount of interface traps. Then we have the dominance of drift current and above-threshold MOSFET operation mode

$$I_{off} \cong \frac{W}{2L} q \frac{\mu_0}{1-\theta V_T} \left(\frac{T_0}{T}\right)^{3/2} \frac{C_{ox} V_T^2(D)}{1 + C_D/C_{ox}}, \quad (14)$$

where the negative threshold voltage plays a role of the effective gate voltage overdrive. The off-state mobile charge density only slightly increases with temperature in this case and, therefore, the leakage current decreases with temperature due to the temperature dependence of carrier mobility.

It is important that the on-state current $I_{on}$ can be simulated on an equal footing with $I_{off}$, using the same relation (3) with $n_{S0} = n_{S0}(V_G = V_{DD})$. The carrier speed saturation effect for short-channel transistors is, of course, relevant in this case. Notice that the current $I_{on}$ in the n-MOSFETs degrades (i.e., decreases) under irradiation on condition of the rebound effect.

### B. Rebound effect and on-state n-MOSFET current compact modeling

The rebound of threshold voltage shift to the positive values at long-term irradiation is a non-trivial property of the n-MOSFET radiation response, which is caused by excess negative interface trap charge [27]. It was commonly believed that due to thin gate oxides, the rebound would become less and less a problem for most modern MOS microelectronics technologies [28]. Nevertheless, the rebound remains relevant at high total doses even for highly scaled circuits [29]. This makes relevant the compact modeling of the rebound effects. The threshold voltage shift can be represented as a sum of shifts due to oxide trap and interface trap buildups [30]

$$\Delta V_T(D) = \Delta V_{ot}(D) + \Delta V_{it}(D) \quad (15)$$

The oxide trap voltage shift can be calculated as follows [31, 32]

$$\Delta V_{ot}(D) = -\frac{qF_{ot}A_d}{C_{ox}} D \left(1 - \frac{\lambda}{\ell}\ln\left(\frac{D}{Pt_{min}}\right)\right) \quad (16)$$

where $P$ is a dose rate, $F_{ot}$ is the deep oxide trap buildup efficiency, $A_d = K_g t_{ox} \eta(E_{ox})$, $\eta(E_{ox})$ is the oxide electric field dependent electron-hole charge yield, $t_{ox}$ is the oxide thickness, $K_g \cong 8\times 10^{12}$ cm$^{-3}$rad(SiO$_2$)$^{-1}$ is the electron-hole pair genera-tion rate constant in SiO$_2$, $\ell$ is the effective width of the oxygen vacancy precursors for the oxide hole traps, $\lambda$ is the minimum tunnel length ($\leq 0.1$ nm). The reference minimum time $t_{min}$ is correlated with $F_{ot}$ and $\ell$ [32] and also depends on temperature [33].

We assume that the interface trap buildup efficiency $F_{it}$ is linearly correlated with the oxide trap efficiency $F_{it} = \beta F_{ot}$, where $\beta$ is assumed to be a dimensionless factor $0 < \beta < 1$. This assumption is reasonable for a long-term irradiation when the retardation of the interface trap buildup can be neglected. Thus, the interface trap voltage shift can be approximated as follows [34]

$$C_{ox}\Delta V_{it}(t) = \beta q F_{ot} A_d D. \qquad (17)$$

Then, the dose dependence of the threshold voltage shift may be written in a concise form

$$C_{ox}\Delta V_T(D) \cong -qF_{ot}A_dD\left(1-\frac{\lambda}{\ell}\ln\left(\frac{D}{Pt_{min}}\right)\right) + \beta qF_{ot}A_dD = \\ = \frac{\lambda}{\ell}F_{ot}A_dD\ln\left(\frac{D}{D_R}\right). \qquad (18)$$

The rebound dose $D_R$, defined as $\Delta V_T(D_R) = 0$, is derived from (18) as follows

$$D_R = Pt_{min}\exp\left[\frac{\ell}{\lambda}(1-\beta)\right] = Pt_{min}\left(\frac{t_{max}}{t_{min}}\right)^{1-\beta}, \qquad (19)$$

where $t_{max} = t_{min}\exp(\ell/\lambda)$ is the maximum tunneling time.

The threshold voltage shift is negative at doses less than $D_R$ due to the dominance of oxide trapped charge and positive at $D > D_R$ due to interface trap prevalence. The rebound dose can be reduced greatly at elevated irradiation temperature [26, 29], likely due to temperature dependence of $t_{min}$.

Figure 6 shows the dose dependencies for the voltage shift and normalized current shifts, simulated for the same 60 nm node MOSFETs as in Fig. 3. The rebound model parameters are fitted to match the extracted $\Delta V_T$ approximately as in Fig. 3. As can be seen in Fig. 6, the simulation results are in good agreement with the experimental data, presented in [5, 29]. The detailed results of the same calculations are plotted in Fig. 7.

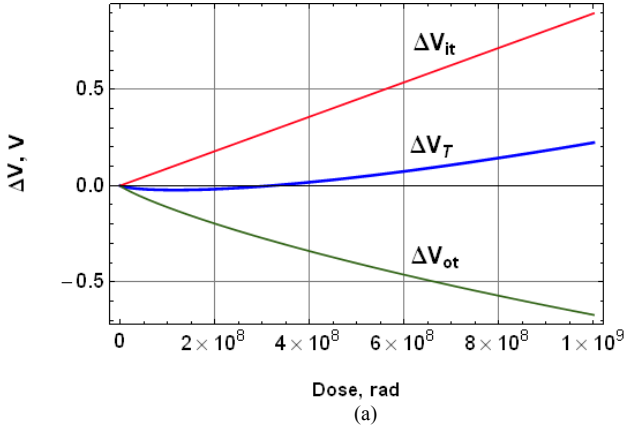

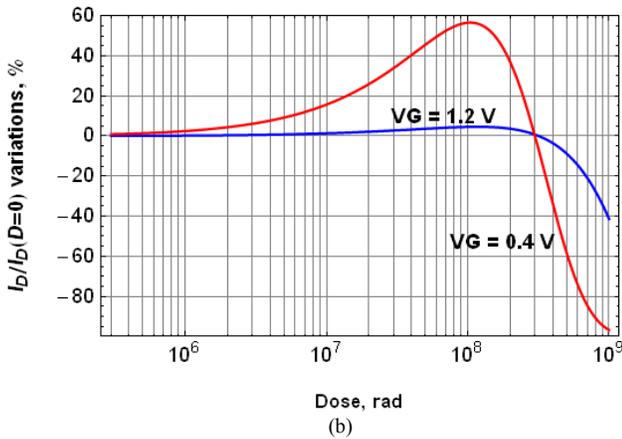

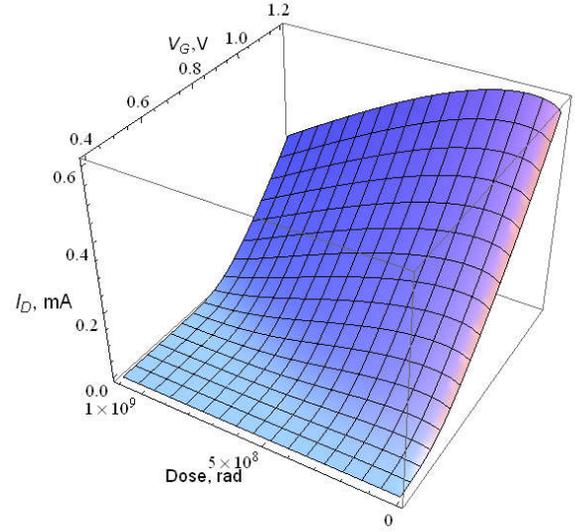

Fig. 7. 3D graphics depicted the dependence of the drain current of the 1 μm/60 nm n-MOSFET as a function of total dose and gate voltage.

Figure 7 clearly exemplifies the n-MOSFET drain current as a function of dose and gate voltage. Having information on the dose behavior of the model parameters, the drain current can be described in a continuous manner over the entire range of radiation and device characteristics. Unfortunately, the radiation and device parameters cannot be uniquely fitted, especially on a restricted set of experimental data. A unique and accurate extraction of the model parameters requires more extended electrical and radiation examination.

Fig. 6. The dose dependencies for the voltage shifts (a) and the normalized drain current variations (b), calculated at $V_G = V_D = 1.2$ V and $V_G = 0.4$ V for the n-MOSFETs with W/L= 1 μm/60 nm, $t_{ox}$ = 3 nm oxide, irradiated with a dose rate $P = 2.5$ krad (Si)/s [5]. Fitted parameters: $F_{ot} = 0.025$, $\beta = 0.11$, $\eta = 1$, $t_{min}=10^{-11}$ s, $\lambda/\ell = 0.024$.

## V. Conclusion

We have presented a concise description of the compact MOSFET physics-based model, which allows us to simulate the MOSFET I-V characteristics both for the long-channel and short-channel MOSFETs over a range of measurement temperatures. It has been shown that the model is suitable for accurate simulation of the MOSFET's I-V characteristics over a wide operation temperature range in harsh environments such as ionizing irradiation with extremely high doses. The model has a closed analytical form with a restricted number of physical parameters, compatible with a standard set of BSIM and SPICE parameters, allowing easy embedding into standard CAD tools via implementation in Verilog-A.